\documentclass[10pt,final,twocolumn]{IEEEtran}

\usepackage[english]{babel}

\usepackage[margin=0.95in]{geometry}
\usepackage{cite}
\usepackage{algpseudocode}
\usepackage{graphicx}
\usepackage{amsmath}
\usepackage{algpseudocode}
\usepackage{algorithm}
\usepackage{caption}
\usepackage{relsize}
\usepackage{subcaption}
\usepackage{amssymb}
\usepackage{float}
\usepackage{ifthen}
\usepackage{multicol}
\usepackage{optidef}
\usepackage{amsmath}
\usepackage{comment}
\usepackage{xcolor}
\usepackage{multirow}
\usepackage{hyperref}

\usepackage{placeins}

\usepackage{caption}

\renewcommand{\baselinestretch}{1.09}

\begin{document}

\raggedbottom

\title{\LARGE{DeepOFW: Deep Learning–Driven OFDM-Flexible Waveform Modulation for Peak-to-Average Power Ratio Reduction}}

\author{Ran Greidi, Kobi Cohen (\emph{Senior Member, IEEE})
    \thanks{The authors are with the School of Electrical and Computer Engineering, Ben-Gurion University of the Negev, Beer Sheva 8410501 Israel. Email: rangrei@post.bgu.ac.il, yakovsec@bgu.ac.il}
    \thanks{This work was supported by the US-Israel Binational Science Foundation (grant No. 2024611).}
    \thanks{This work has been submitted to the IEEE for possible publication.
    Copyright may be transferred without notice, after which this version may
    no longer be accessible.}
	\vspace{-0.75cm}
}

\date{2026}


\maketitle
\begin{abstract}
Peak-to-average power ratio (PAPR) remains a major limitation of multicarrier modulation schemes such as orthogonal frequency-division multiplexing (OFDM), reducing power amplifier efficiency and limiting practical transmit power. In this work, we propose DeepOFW, a deep learning–driven OFDM-flexible waveform modulation framework that enables data-driven waveform design while preserving the low-complexity hardware structure of conventional transceivers. The proposed architecture is fully differentiable, allowing end-to-end optimization of waveform generation and receiver processing under practical physical constraints. Unlike neural transceiver approaches that require deep learning inference at both ends of the link, DeepOFW confines the learning stage to an offline or centralized unit, enabling deployment on standard transmitter and receiver hardware without additional computational overhead. The framework jointly optimizes waveform representations and detection parameters while explicitly incorporating PAPR constraints during training. Extensive simulations over 3GPP multipath channels demonstrate that the learned waveforms significantly reduce PAPR compared with classical OFDM while simultaneously improving bit error rate (BER) performance relative to state-of-the-art transmission schemes. These results highlight the potential of data-driven waveform design to enhance multicarrier communication systems while maintaining hardware-efficient implementations. An open-source implementation of the proposed framework is released to facilitate reproducible research and practical adoption.
\end{abstract}

\begin{IEEEkeywords}
End-to-end (E2E) learning, multicarrier modulation, peak-to-average power ratio (PAPR), learned waveform design, OFDM-based systems.
\end{IEEEkeywords}

\section{Introduction}
\label{Introduction}

Orthogonal frequency-division multiplexing (OFDM) is widely used in modern wireless communication systems due to its robustness to multipath propagation and its low-complexity implementation, and continues to serve as the foundation for many advanced waveform designs investigated for future wireless systems, including 6G and beyond \cite{juwono2021future}. Despite these advantages, OFDM exhibits several inherent limitations, including high peak-to-average power ratio (PAPR) and significant out-of-band emissions. These drawbacks become increasingly problematic in emerging communication scenarios such as high-frequency operation, energy-constrained devices, and dense network deployments envisioned for beyond-5G systems. As a result, numerous alternative waveform designs and OFDM-based variants have been proposed, including single-carrier frequency-division multiple access (SC-FDMA) \cite{farhang2024sc}, which targets PAPR reduction, as well as orthogonal time–frequency space (OTFS) modulation~\cite{shafie2024coexistence}, generalized frequency division multiplexing (GFDM) \cite{tai2020overview}, orthogonal chirp division multiplexing (OCDM) \cite{ouyang2016orthogonal}, 
orthogonal chirp-frequency division multiplexing (OCFDM) \cite{moreira2024orthogonal}, among others. While these approaches address specific limitations of OFDM, such as improved power efficiency or enhanced robustness to time-varying channels, they often introduce additional signal processing complexity or impose new design trade-offs at the transmitter and receiver. Consequently, mitigating these limitations while preserving the simplicity of conventional transceiver architectures remains an open challenge.

Recent advances in deep learning (DL) have led to the emergence of neural transceivers, in which the transmitter and receiver are jointly optimized in an end-to-end (E2E) manner\cite{oshea2017introduction, aoudia2018end, qin2019deep}. This approach have demonstrated the potential of data-driven design to move beyond the limitations of conventional communication architectures, particularly in challenging propagation environments or under strict system constraints. At the same time, the practical deployment of fully neural transceivers remains an open challenge. Implementing deep neural networks at both ends of the link entails considerable computational complexity, power consumption, and latency, which are difficult to resolve with the requirements of low-cost, low-power devices. These limitations are particularly critical in the context of emerging 6G use cases, such as massive Internet-of-Things (IoT) deployments and ultra-low-power edge devices, where real-time DL inference at high data rates is difficult to sustain with current hardware capabilities. These considerations motivate the search for learning-based communication paradigms that retain the performance benefits of data driven E2E optimization, while preserving the simplicity and efficiency of conventional transceiver architectures.

\subsection{Main Results}

Seeking to close the implementation gap between DL-driven PHY and current hardware complexity, this work proposes a paradigm shift that decouples the computational intensity of deep learning from the physical-layer implementation. To address this challenge, we suggest localizing the DL inference stage at the Access Point (AP) or a centralized computing unit, thereby enabling the use of optimized, data-driven waveforms without requiring complex neural processing at the terminal side. Building on this idea, we propose DeepOFW-modulation, a novel hardware-efficient PHY architecture designed to replicate the low-complexity footprint of current transceiver hardware. Unlike traditional modular designs, the proposed DeepOFW-modulation architecture is fully differentiable, allowing it to be integrated into a comprehensive E2E learning framework. This differentiability enables joint optimization of the transmit signal and receiver processing via direct backpropagation while simultaneously satisfying practical system constraints. Ultimately, when optimized through our data-driven framework, DeepOFW-modulation achieves superior reliability and power efficiency compared with conventional state-of-the-art transmission schemes, while maintaining hardware complexity comparable to current-generation stations.

The main contributions of this paper are summarized as follows:

\noindent
\textbf{1) Novel Differentiable PHY Architecture:} We introduce DeepOFW-modulation, a hardware-efficient physical-layer architecture designed to be fully differentiable. Unlike conventional modular designs, this architecture allows for the seamless integration of hardware constraints into a gradient-based optimization loop, enabling direct E2E training of the communication chain.

\noindent
\textbf{2) Centralized Learning with Hardware-Efficient Terminals:} We propose a communication paradigm in which the DL inference stage is centralized at the AP or a dedicated computing unit. By leveraging the differentiability of the DeepOFW-modulation architecture during an offline training phase, the system learns optimized signaling strategies that can be executed by the AP in real time. This design enables terminal devices to operate using conventional low-complexity hardware, as they interact with the learned waveforms without requiring on-board neural processing or additional computational overhead.

\noindent
\textbf{3) Joint Optimization under Physical Constraints:} We propose an E2E learning framework that leverages the differentiability of DeepOFW-modulation. Our framework enables the joint optimization of the transmitter and receiver while incorporating explicit learned PAPR constraint tailored specifically to the channel condition. This ensures that the learned waveforms achieve strong error-rate performance while remaining practical for power-limited and nonlinear hardware environments.

\noindent
\textbf{4) Performance Validation:} Through extensive simulations, we demonstrate that the proposed DeepOFW-modulation architecture, when optimized via our E2E framework, significantly outperforms conventional benchmarks. We show improvements in both BER and power efficiency across various challenging channel conditions, without increasing the hardware complexity.

\noindent
\textbf{5) Open source software:} We develop and release an open-source implementation of the proposed DeepOFW framework. The implementation is built using Sionna~\cite{sionna} and is publicly available on GitHub (see \cite{deepofw_code}). Making the framework publicly accessible promotes reproducible research and enables other researchers and practitioners to explore, extend, and adapt DeepOFW for a wide range of wireless communication scenarios.

We based our formulation upon a generalized framework of advanced multicarrier and transform-domain modulation schemes such as OFDM, OTFS, SC-FDMA and more, leveraging a learned multi waveforms modulation. The learned waveforms optimized with a PAPR constrain, are used to form the transmitted signal. At the receiver side, the same waveforms are used together with a learned one tap equalizer to recover the transmitted bits. No DL infrastructure is needed at the transmitter or receiver side. Our approach is evaluated on the common 3GPP multi-path channel, showing its ability to adapt its waveforms according to the channel used so PAPR is reduced while transmission rate is maximized. Our results analysis shows that the generated waveforms are showing to adjust their division between time to frequency according to the channel used, so that PAPR is optimized. In a low delay spread channel, the waveforms spanned across frequency, while in the high delay spread cases, they spanned across time.

To the best of our knowledge, this is the first work to propose a fully differentiable PHY architecture that enables E2E data-driven optimization while confining the DL inference stage entirely to a centralized unit. Unlike existing neural transceivers that require non-standard processing at both ends of the link, DeepOFW-modulation combines the performance benefits of learned waveforms with the hardware constraints of current-generation transceivers.

\subsection{Related Work}
\label{Related Work}

A wide range of waveform design and spectrum shaping techniques have been proposed to overcome the limitations of conventional OFDM, particularly its high PAPR and sensitivity to channel impairments. Early works have explored parametric optimization of transmit spectra in communication systems~\cite{naparstek2012parametric}. SC-FDMA was introduced as a DFT-spread variant of OFDM to reduce PAPR and improve power efficiency, and has been adopted in uplink transmission standards~\cite{Holma2009}. More recently, alternative modulation schemes such as OTFS, GFDM and OCDM have been investigated to improve robustness in time-varying or highly frequency-selective channels~\cite{9661102,Matthé2016,9013425}. While these approaches address specific shortcomings of OFDM, they typically rely on more complex signal processing, increased receiver complexity, asymmetric burden between transmitter and receiver or additional system overhead.

More recently, E2E learning approaches have been proposed as an alternative paradigm for communication system design, in which the transmitter and receiver are jointly optimized using neural networks. By modeling the entire communication chain as an autoencoder (AE), these methods enable data-driven optimization of modulation, coding, and detection, often achieving performance gains in challenging or highly nonlinear channel conditions~\cite{9905727,10589475,9446711,10485272,11053759}. 

Recent E2E learning approaches can generally be categorized into two classes: fully AE-based transceivers, and OFDM-based AE frameworks. In the former, both the transmitter and receiver are entirely replaced by neural network–based encoder and decoder blocks. For example, in~\cite{9747919}, an E2E learning framework has been proposed for waveform design in which the transmit filter, constellation, and receiver are jointly optimized under explicit constraints on PAPR and adjacent channel leakage. The proposed approach preserves a conventional single-carrier transmission structure while leveraging neural networks at the receiver to maximize an achievable information rate. Simulation results demonstrate improved spectral containment and power efficiency compared to conventional OFDM, without increasing transmitter complexity, but still requires a heavy deep neural receiver to account for the channel. A related line of work is presented in~\cite{9745361}, where a model-based E2E learning framework is developed for wavelength-division multiplexing (WDM) systems with transceiver hardware impairments. In this approach, physical-layer knowledge is incorporated into the learning process to jointly optimize pulse shaping and symbol mapping under practical hardware constraints. While this method demonstrates improved spectral efficiency and robustness to hardware impairments, it similarly relies on neural processing at the receiver and is primarily tailored to optical communication scenarios. For the later, AE-based methods have been integrated with OFDM systems. In~\cite{9271932}, the authors combine an AE with a channel-coded OFDM framework to jointly learn adaptive modulation and coding strategies under varying channel conditions.

In parallel with advances in learning-based waveform design, learning-based methods have also been widely investigated for spectrum allocation and channel access among multiple users in wireless networks. In these works, the wireless environment is often modeled as a sequential decision-making problem where users learn channel availability or quality over time. Distributed optimization~\cite{leshem2012multichannel, gafni2022distributed, ami2024stable}, multi-armed bandit~\cite{Tekin_2012_Online, liu2012learning, cohen2014restless, bistritz2018distributed, gafni2021learning, jiang2023online, raman2024global} and deep reinforcement learning~\cite{naparstek2018deep, liu2021dynamic, bokobza2023deep, paul2023multi, cohen2024sinr} frameworks have been applied to problems such as opportunistic spectrum access, distributed channel selection, and dynamic spectrum sharing. These approaches focus primarily on learning channel statistics or access policies while assuming conventional physical-layer transmission schemes.

State-of-the-art E2E transceiver designs often rely on deep neural networks at both the transmitter and receiver, resulting in high computational complexity, increased latency, and limited practicality for real-world deployment, particularly in power and cost-constrained devices. This limitation motivates the development of learning-based frameworks that retain the benefits of E2E optimization while preserving the structure and efficiency of conventional waveform designs.

The reminder of this paper is organized as follows: Section II establishes the system model and problem statement, while Section III details our proposed modulation architecture. Section IV describes the suggested learning framework and finally, performance analysis and concluding remarks are provided in Sections V and VI, respectively.\vspace{0.2cm}

\noindent
\textbf{Notations:} Lowercase and uppercase bold letters denote vectors and matrices, respectively, while regular italic letters represent scalars. The superscripts $(\cdot)^T$, $(\cdot)^H$, and $(\cdot)^*$ denote the transpose, Hermitian transpose, and complex conjugate, respectively. The imaginary unit is denoted by $j = \sqrt{-1}$. The expectation operator is written as $\mathbb{E}[\cdot]$, and $\|\cdot\|$ denotes the Euclidean norm. Continuous-time convolution is represented by the operator $\ast$. The sets of real and complex numbers are denoted by $\mathbb{R}$ and $\mathbb{C}$, respectively.

Table~\ref{table:notation} summarizes the notation used throughout this paper.

    \begin{table} [htbp!]
    \centering
    \caption{Summary of Notation}
    
    \label{table:notation}
    \begin{center}
    \begin{tabular}{ |p{1cm}||p{4.5cm}|}
    \hline
     Notation& Description \\
    \hline \hline
    $N$ & Number of waveforms used in a single DeepOFW symbol\\ 
    \hline
    $M$ & Number of bits per symbol \\ 
    \hline     
    $B_{n,m}$ & $m$th bit from constellation symbol $n$ \\ 
    \hline    
    $f_k(\cdot)$ & $k$th waveform function \\ 
    \hline      
    $T$ & Transmit block symbol length \\ 
    \hline
    $T_g$ & Transmit block guard interval length \\ 
    \hline
    $L$ & Channel multi path reflections \\ 
    \hline    
    $S$ & Batch size \\ 
    \hline 
    $\mathbf{x}$ &  Symbol vector\\ 
    \hline     
    $\tilde{\mathbf{x}}$ &  Recovered symbol vector\\ 
    \hline 
    $\mathbf{r}$ &  Match filter output, the detector input\\ 
    \hline 
    $\mathbf{q}$ &  Detector parametrization vector\\ 
    \hline 
    $\mathbf{Q}$ &  Learned waveforms matrix\\ 
    \hline 
    $D_{\mathbf{q}}\{ \cdot\}$ &  Detector parametrized with a vector $\mathbf{q}$\\ 
    \hline
    
    \end{tabular}
    \end{center}  
    \end{table}

\section{System Model and Design Objective}
\label{sec:system_model}

We consider a multicarrier wireless communication system operating over $N$ parallel subchannels, following the general principles of OFDM-type modulation. In such systems, information bits are mapped to complex-valued symbols that are transmitted simultaneously across multiple frequency resources and combined to form a time-domain signal. After propagation through the wireless channel, the receiver processes the signal to recover the transmitted symbols.

Let $\mathbf{B} \in \{0,1\}^{N \times M}$ denote the matrix of bits within a transmit block, consisting of $N$ baseband symbols each carrying $M$ bits. Using a bit-interleaved coded modulation (BICM) scheme, the bits are mapped to constellation symbols $\mathbf{x} \in \mathbb{C}^N$ drawn from a constellation set $\mathcal{C} \subset \mathbb{C}$, for example a $2^M$-QAM constellation with Gray labeling.

We consider an OFDM-type multicarrier transmission model, where the symbol vector $\mathbf{x}$ is mapped onto orthogonal basis functions through the inverse discrete Fourier transform (IDFT).

To overcome the limitations associated with the high PAPR of OFDM signals, as well as the complexity trade-offs introduced by existing mitigation techniques, , discussed in Section~\ref{Introduction}, we develop a generalized multicarrier modulation framework that reduces PAPR while preserving the low-complexity hardware structure of OFDM-type transceivers. 

In particular, we aim to maintain transmitter and receiver architectures based on simple linear transforms and lightweight signal processing operations, enabling practical deployment on existing communication hardware. By ensuring that the transformations used in the proposed architecture are fully differentiable, we facilitate the use of E2E optimization to jointly refine the transceiver chain without sacrificing its streamlined implementation.

\section{DeepOFW Modulation Architecture} 
\label{System Model and Problem Statement}

This section introduces the proposed DeepOFW modulation architecture. The framework generalizes classical OFDM by replacing the fixed Fourier waveform basis with a learnable waveform set that can adapt to the wireless channel conditions. The architecture is designed to remain fully compatible with conventional transmitter and receiver hardware while enabling data-driven optimization through a differentiable system formulation. We first present the continuous-time signal model and its discrete implementation, showing how classical OFDM arises as a particular instance of the proposed framework. We then discuss practical deployment considerations for integrating DeepOFW into real communication systems. The learning framework used to optimize the waveform representations and detection parameters is described in the following section.

\begin{figure*}[t]
    \centering

    \begin{subfigure}[t!]{0.85\textwidth}%
        \centering
        \includegraphics[width=\textwidth]{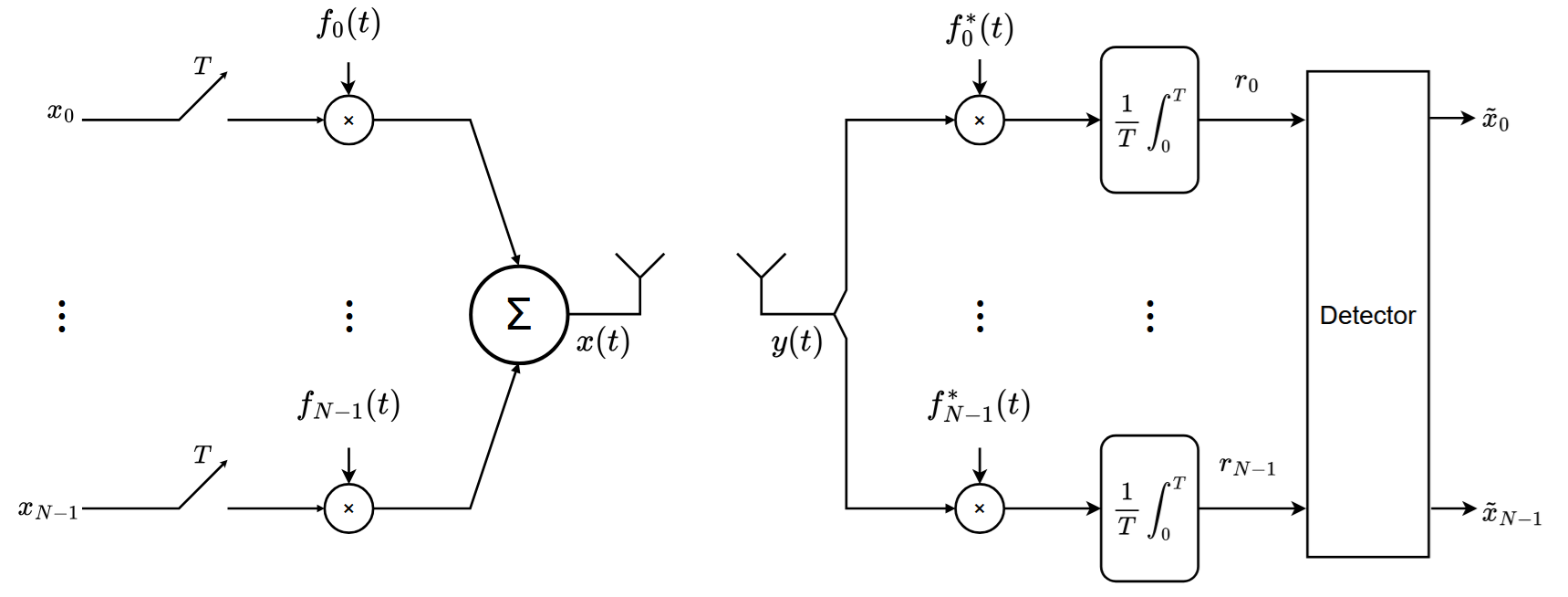}
        \caption{Continuous-time representation of the considered system model.}
        \label{fig:system figure analoge}
    \end{subfigure}\par\medskip

    \vspace{3mm}
    
    \begin{subfigure}[b]{0.85\textwidth}%
        \centering
        \includegraphics[width=\textwidth]{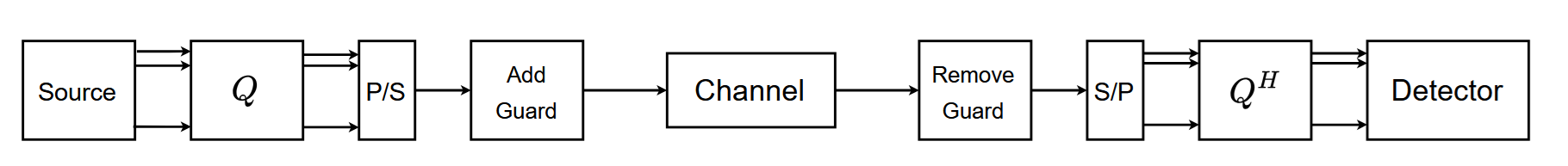}
        \caption{Discrete-time implementation of the system model.}
        \label{fig:system figure discrete}
    \end{subfigure}\par\medskip

    \caption{Illustration of the proposed OFDM-type transmission framework in continuous-time and discrete-time implementations.}
    \label{fig:system_model}
\end{figure*}

\subsection{DeepOFW Signal Model}

We consider an OFDM-type multicarrier transmission system in which the transmitted signal is constructed using a set of waveform functions $\{f_k(t)\}_{k=0}^{N-1}$. Let $\mathbf{B} \in \{0,1\}^{N \times M}$ denote the matrix of transmitted bits. Using a BICM scheme, the bits are mapped to constellation symbols $\mathbf{x} \in \mathbb{C}^N$ drawn from a constellation set $\mathcal{C}$.

A cyclic prefix of duration $T_g$ is introduced to mitigate inter-symbol interference. The resulting time-domain transmitted signal, illustrated in Fig.~\ref{fig:system_model}, is given by

\begin{equation}
\label{Tx time domain}
x(t) = W_{[-T_g,T]}(t) \sum_{k=0}^{N-1} x_k f_k(t),
\end{equation}

where $W_{[-T_g,T]}(t)$ denotes a pulse-shaping or windowing function that is zero outside the interval $[-T_g,T]$.

The wireless channel is modeled as a multipath channel with impulse response

\begin{equation}
\label{channel transfer function}
h(\tau) = \sum_{l=0}^{L-1} a_l \delta(\tau-\tau_l),
\end{equation}

where $L$ is the number of paths, and $a_l$ and $\tau_l$ denote the complex gain and delay of the $l$-th path.

The received signal is therefore

\begin{equation}
\label{channel out signal}
y(t) = (h*x)(t) + n(t),
\end{equation}
where $n(t)$ denotes additive white Gaussian noise.

After removing the cyclic prefix, the receiver performs matched filtering using the waveform functions. The projection onto the $n$-th waveform yields

\begin{equation}
r_n = \frac{1}{T}\int_0^T y(t)f_n^*(t)dt.
\end{equation}
This leads to
\begin{equation}
r_n = \sum_{k=0}^{N-1} x_k \Lambda_n[k] + \tilde{n}_n,
\end{equation}
where
\[
\Lambda_n[k] = \sum_{l=0}^{L-1} a_l \frac{1}{T} \int_0^T f_k(t-\tau_l)f_n^*(t)dt.
\]
The recovered symbol is obtained through a detector
\[
\tilde{x}_n = D\{r_n\}.
\]

This formulation describes a generalized multicarrier system for arbitrary waveform sets $\{f_k(t)\}_{k=0}^{N-1}$.

\subsection{Discrete-Time Implementation}

In practical implementations, OFDM systems operate in discrete time using DFT/IDFT transforms. The continuous-time model can therefore be represented using sampled waveform functions

\[
f_k[n] = f_k\!\left(\frac{nT}{N}\right), \quad n=0,\ldots,N-1.
\]

Stacking these samples forms a waveform matrix $\mathbf{Q} \in \mathbb{C}^{N\times N}$ whose $k$-th column corresponds to waveform $\mathbf{f}_k$.

The transmitted signal can then be written as

\[
\mathbf{x}_t = \mathbf{Q}\mathbf{x}.
\]

At the receiver, matched filtering is performed via multiplication with $\mathbf{Q}^H$, followed by a detector that estimates the transmitted symbols.

Classical OFDM is recovered as a special case when $\mathbf{Q}$ is chosen as the IFFT matrix and a one-tap equalizer is applied and serves as the detector.

This observation highlights the generality of the DeepOFW framework: OFDM corresponds to a particular choice of waveform matrix $\mathbf{Q}$, while alternative waveform bases can be incorporated within the same architecture to better adapt to channel conditions or optimization objectives.

\subsection{Practical Deployment of DeepOFW}

The goal of DeepOFW is to enable data-driven waveform optimization while preserving the low-complexity hardware structure of conventional transceivers. Specifically, the system seeks to jointly optimize the waveform matrix $\mathbf{Q}$ and the detector parameters $\mathbf{q}$ associated with a practical detection operator $D_{\mathbf{q}}\{\cdot\}$.

Importantly, DeepOFW does not require deep neural network inference at the transmitter or receiver. Instead, the learning process is performed offline or at a centralized network entity such as an access point (AP). The resulting waveform matrix $\mathbf{Q}$ and detector parameters $\mathbf{q}$ are then distributed to the stations.

\begin{figure}[htp]
    \centering
    \includegraphics[width=8cm]{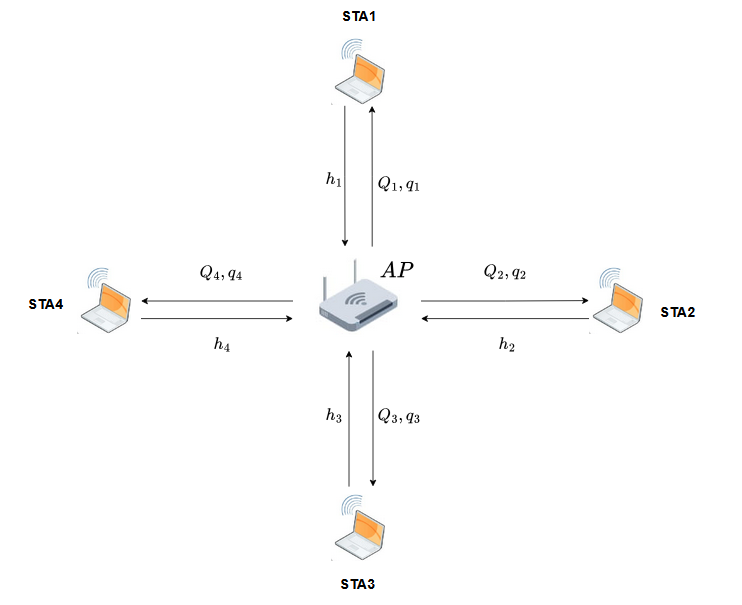}    \caption{Possible deployment strategy of DeepOFW in a star topology, suitable for WLAN 802.11 standards. Each station interacts with the AP to obtain $\mathbf{q}$ and $\mathbf{Q}$. The channel state information is first delivered to the AP by the stations. The AP, equipment with DL accelerators, generate $\mathbf{q}_i,\mathbf{Q}_i$ for station $i$ accordingly, and and communicates them back to the station.}
    \label{fig:star topology workflow ilustration}
\end{figure}

This centralized learning paradigm is particularly suitable for WLAN systems. In IEEE 802.11 networks, multiple stations communicate through an access point that coordinates transmissions. The AP can therefore estimate channel conditions, compute optimized waveform parameters $(\mathbf{Q}_i,\mathbf{q}_i)$ for each station $i$, and distribute them during the handshake or preamble phase, as illustrated in Fig~\ref{fig:star topology workflow ilustration}. 

As channel conditions evolve, the AP can update these parameters and redistribute them to the stations. This approach enables adaptive waveform optimization while maintaining simple station-side hardware consisting primarily of matrix multiplications and lightweight DSP operations.

\section{DeepOFW Learning Framework}
\label{qQ Waveforms Generation}

Building upon the DeepOFW architecture described in the previous section, we now address the problem of generating the waveform matrix and associated detector parameters that define the modulation scheme. In particular, our goal is to determine the waveform basis $\mathbf{Q}$ and detector parameters $\mathbf{q}$ that jointly optimize the system performance under practical communication constraints.

To this end, we formulate a data-driven optimization framework that leverages E2E learning to design the DeepOFW waveforms. The optimization problem jointly considers the waveform generation, the detector implementation, and the system objective. Importantly, the resulting transceiver structure remains compatible with simple hardware implementations, as it relies only on linear transforms and lightweight digital signal processing modules.

In the following subsections, we describe the detector implementation, the optimization objective, and the training architecture used to learn the DeepOFW waveforms.

\subsection{Detector implementation}

\begin{figure}[htp]
    \centering
    \includegraphics[width=4cm]{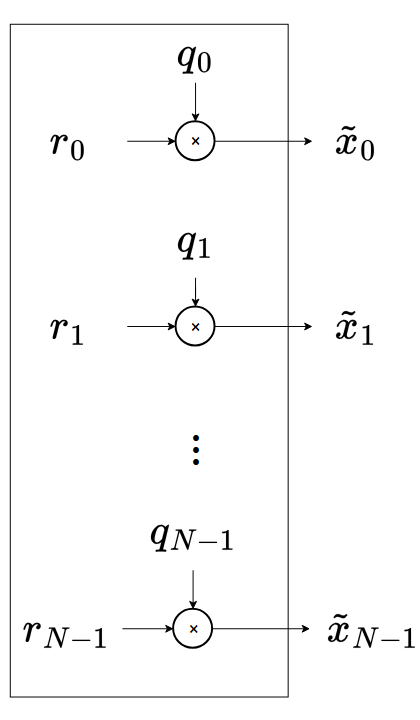}    \caption{An illustration of the element-wise multiplication detector $D_{\mathbf{q}}\{ \mathbf{r}\}$.}
    \label{fig:detctor}
\end{figure}

The receiver employs a one-tap detector operating through element-wise vector multiplication. The detected constellation symbol is given by

\begin{equation}
\label{one tap detector}
\tilde{x}_n = D_{\mathbf{q}}\{r_n\} = r_n q_n,
\end{equation}
where $r_n$ denotes the matched-filter output corresponding to the $n$-th waveform and $q_n$ is the learned detector coefficient. The vector $\mathbf{q}$ therefore defines the equalization parameters applied to the received signal. The detector structure adopted in this work is illustrated in Fig.~\ref{fig:detctor}.

\subsection{Learning Objective and PAPR-Constrained Optimization}

The DeepOFW framework is trained using an E2E learning objective that jointly considers symbol detection accuracy and PAPR reduction. In typical E2E communication systems, the training objective aims to maximize the achievable rate or, equivalently, minimize the binary cross-entropy (BCE) between the transmitted bits and the receiver's posterior estimates obtained from the detected constellation symbols~\cite{9508784}. Following this approach, we define the first objective using the posterior distribution of the transmitted bits derived from the log-likelihood ratios (LLRs).

Let $P_{\mathbf{Q},\mathbf{q}}(b_{n,m}|\tilde{x})$ denote the posterior probability of the $m$-th bit of the $n$-th recovered constellation symbol obtained from the detected symbol $\tilde{x}$ under waveform matrix $\mathbf{Q}$ and detector parameters $\mathbf{q}$. Assuming a batch size of $S$ and model parameters $\mathbf{\theta}$, the BCE loss is given by

\begin{equation}
\label{BCE loss objective_1}
\begin{multlined}
\mathcal{R}(\mathbf{B};\mathbf{\theta}) =
-\frac{1}{NM}\sum_{n=0}^{N-1}\sum_{m=0}^{M-1}
\mathbb{E}\!\left[\log_2\big(P_{\mathbf{Q},\mathbf{q}}(B_{n,m}|\tilde{\mathbf{x}})\big)\right]
\vspace{0.2cm}\\
\approx
-\frac{1}{NMS}\sum_{s=0}^{S-1}\sum_{n=0}^{N-1}\sum_{m=0}^{M-1}
\log_2\!\big(P_{\mathbf{Q}^s,\mathbf{q}^s}(B_{n,m}^s|\tilde{\mathbf{x}}^s)\big).
\end{multlined}
\end{equation}

In addition to detection performance, the DeepOFW framework explicitly incorporates a PAPR constraint during training. Let $p_{\mathbf{\theta}}[n]$ denote the instantaneous transmit power at time sample $n$,
$p_{\mathbf{\theta}}[n] = |(\mathbf{Q}\mathbf{x})_n|^2$, and let $\bar{p}_{\mathbf{\theta}}$ denote the average transmit power. PAPR reduction can be formulated by minimizing the probability that the instantaneous power exceeds a threshold. This is equivalent to minimizing
$\mathbb{E}\!\left[\max\!\left(\frac{p(t)}{\bar{p}}-\nu,0\right)\right]$ for $t \sim U[-\frac{T}{2},\frac{T}{2}]$~\cite{9747919}. Based on this formulation, we define the empirical PAPR loss as

\begin{equation}
\label{PAR loss objective_emprical}
\begin{multlined}
\mathcal{P}(\mathbf{\theta}) =
\mathbb{E}\!\left[\max\!\left(\frac{p_{\mathbf{\theta}}[n]}{\bar{p}_{\mathbf{\theta}}}-\epsilon_{\mathbf{\theta}},0\right)\right]
\\
\approx
\frac{1}{SN}\sum_{s=0}^{S-1}\sum_{n=0}^{N-1}
\max\!\left(\frac{p_{\mathbf{\theta}}^s[n]}{\bar{p}_{\mathbf{\theta}}^s}-\epsilon_{\mathbf{\theta}}^s,0\right).
\end{multlined}
\end{equation}

The parameter $\epsilon_{\theta}$ acts as a learnable PAPR threshold and is jointly optimized with the model parameters. Channels with larger delay spread typically require higher PAPR to achieve reliable detection, whereas channels with smaller delay spread allow stronger PAPR reduction. Allowing $\epsilon_{\theta}$ to adapt during training enables the model to balance waveform complexity and detection performance according to the channel conditions.

In practice, $\epsilon_{\theta}$ is generated by a lightweight neural module that takes the channel realization as input and outputs a scalar value, denoted by $\epsilon_{\theta} = g_{\theta}(h).$ To prevent the model from trivially increasing the PAPR threshold, we introduce an additional regularization term

\begin{equation}
\label{teta loss objective}
\Theta(\mathbf{\theta}) =
\frac{1}{S}\sum_{s=0}^{S-1}\epsilon_{\mathbf{\theta}}^s,
\end{equation}
which encourages the learned threshold to remain as small as possible while maintaining reliable symbol detection.

\subsection{Power Normalization and Constrained Optimization Objective}

In practical communication systems, the transmitted signal must satisfy a power constraint to ensure compatibility with hardware limitations and regulatory requirements~\cite{8054694}. Therefore, the waveform matrix $\mathbf{Q}$ must be normalized so that the average transmit energy remains fixed.

Let the transmitted signal be $\mathbf{Q}\mathbf{x}$, where $\mathbf{x}$ denotes the vector of transmitted constellation symbols. We impose the average energy constraint $\mathbb{E}\!\left[\|\mathbf{Q}\mathbf{x}\|_2^2\right] = N.$

Assuming the transmitted symbols are independent and identically distributed and the constellation is normalized such that $\mathbb{E}[\mathbf{x}\mathbf{x}^H] = \mathbf{I}_{N\times N},$ the expected transmit energy can be written as
\begin{equation}
\label{normalization factor}
\mathbb{E}\!\left[\|\mathbf{Q}\mathbf{x}\|_2^2\right]
= \mathbb{E}\!\left[\mathbf{x}^H \mathbf{Q}^H \mathbf{Q}\mathbf{x}\right] 
= \mathrm{trace}\!\left(\mathbf{Q}\mathbf{Q}^H\right).
\end{equation}

To satisfy the transmit power constraint, the waveform matrix is therefore scaled by the normalization factor $\sqrt{\frac{N}{\mathrm{trace}(\mathbf{Q}\mathbf{Q}^H)}}.$

Combining the detection objective and the PAPR regularization terms introduced earlier, the overall optimization problem becomes

\begin{equation}
\label{total loss objective}
\begin{aligned}
\min_{\theta} \quad & 
\alpha \mathcal{R}(\theta)
+ \beta \mathcal{P}(\theta)
+ \gamma \Theta(\theta) \\
\textrm{s.t.} \quad &
\mathbb{E}\!\left[\|\mathbf{Q}\mathbf{x}\|_2^2\right] = N.
\end{aligned}
\end{equation}



\subsection{Adaptive Multi-Objective Loss Balancing}

The DeepOFW training objective combines multiple loss terms that operate on different numerical scales. To ensure stable optimization when training with stochastic gradient descent (SGD), we develop an uncertainty-based loss weighting mechanism inspired by~\cite{Kendall_2018_CVPR} and adapted to the channel-dependent objectives of the DeepOFW framework. Under this formulation, the contribution of each objective is automatically balanced during training using learnable uncertainty parameters. The overall loss function is therefore defined as

\begin{equation}
\label{total loss objective_plus_uncertainty}
\begin{multlined}
\mathcal{L} =
e^{-\sigma_{\mathcal{R}}(h)} \mathcal{R}
+ e^{-\sigma_{\mathcal{P}}(h)} \mathcal{P}
+ e^{-\sigma_{\Theta}(h)} \Theta \\
+ \sigma_{\mathcal{R}}(h)
+ \sigma_{\mathcal{P}}(h)
+ \sigma_{\Theta}(h).
\end{multlined}
\end{equation}
Here, $\sigma_{\mathcal{R}}(h)$, $\sigma_{\mathcal{P}}(h)$, and $\sigma_{\Theta}(h)$ denote learnable uncertainty parameters conditioned on the channel realization $h$. These parameters enable adaptive scaling of the different loss components during training.

This mechanism allows the DeepOFW framework to dynamically balance competing objectives, such as reliable symbol detection and PAPR reduction, in a data-driven manner while maintaining stable optimization across varying channel conditions.

\begin{figure*}[h!]
    \centering
    \includegraphics[width=\textwidth]{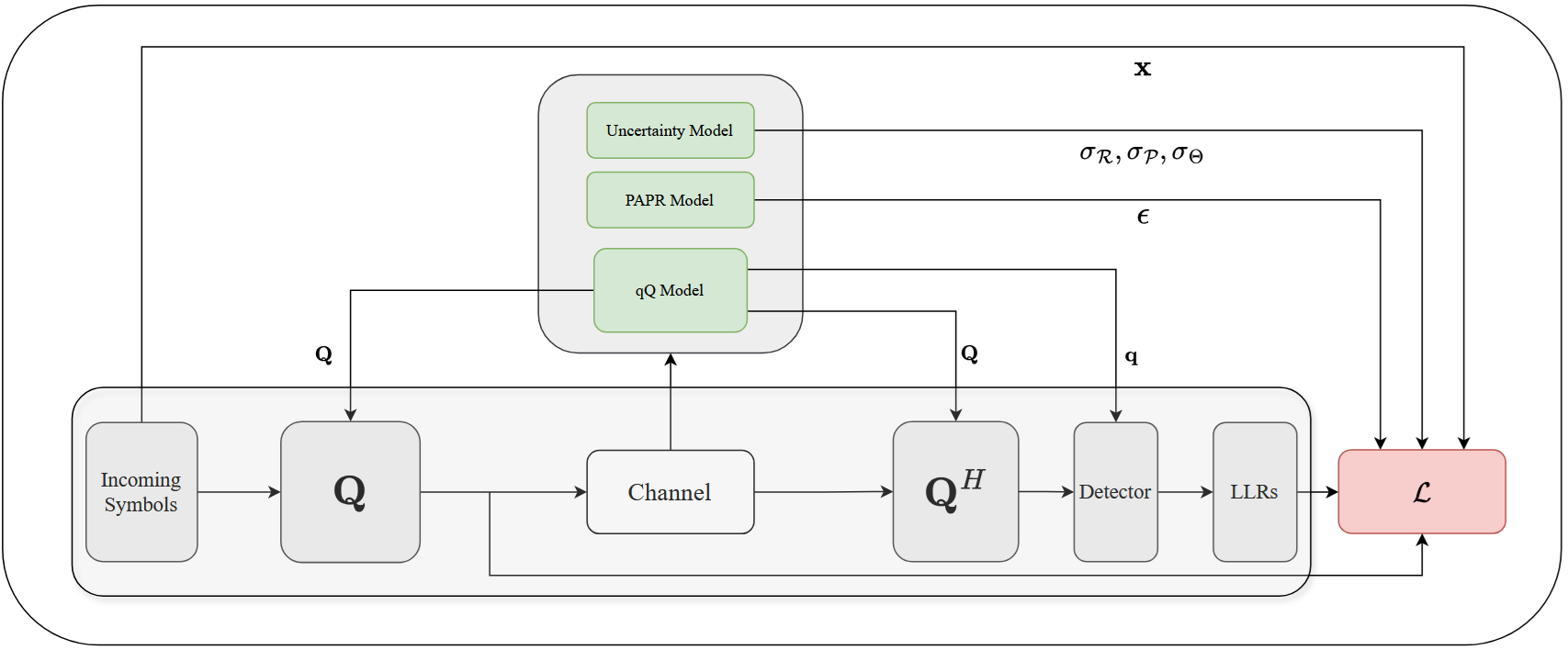}
    \caption{E2E model architecture. The transmitter–receiver chain is trained end-to-end to learn a complex waveform matrix $\mathbf{Q}$ and detector parameters $\mathbf{q}$ from instantaneous time-domain channel impulse responses. A GRU-based network generates $\mathbf{Q}$ and $\mathbf{q}$, which are applied by the $\mathbf{Q}$ modulator and $\mathbf{Q}$ demodulator together with a one-tap detector to produce LLRs for decoding. An uncertainty network and a PAPR network predict per-sample weighting factors and a regularization coefficient, enabling joint optimization of BER and PAPR during training.}
    \label{e2e arch}
\end{figure*}

\subsection{DeepOFW E2E Learning Architecture}

To optimize the DeepOFW system, we implement the entire communication chain as a differentiable E2E model that can be trained using SGD. All signal processing operations, from waveform generation at the transmitter to symbol detection at the receiver, are expressed through differentiable operations, enabling gradient-based optimization of the system parameters. After training, the learned parameters are exported offline to generate the waveform matrix $\mathbf{Q}$ and detector coefficients $\mathbf{q}$ used in the operational communication system.

During training, each sample within a mini-batch is associated with a randomly generated channel realization. Specifically, a time-domain multipath channel is generated according to a randomly selected delay spread together with a noise variance. Independent and identically distributed constellation symbols are also generated for each sample. For every sample $s$ in the batch, the parameters $\{\mathbf{Q}^{s},\mathbf{q}^{s},\sigma_{\mathcal{R}}^{s},\sigma_{\mathcal{P}}^{s},\sigma_{\Theta}^{s},\epsilon^{s}\}$ are produced independently based on the corresponding channel realization.

The padded time-domain channel impulse response serves as the input to several neural modules responsible for generating the DeepOFW parameters. These include the waveform generator producing $\mathbf{Q}$ and $\mathbf{q}$, the uncertainty-weighting model, and the module producing the PAPR threshold parameter $\epsilon_{\theta}$. Using the generated waveform matrix, the transmit signal $\mathbf{Q}\mathbf{x}$ is constructed and extended with a cyclic prefix before passing through the convolutional channel model. At the receiver, the cyclic prefix is removed, followed by matched filtering through multiplication with $\mathbf{Q}^H$. The detector then applies element-wise multiplication with $\mathbf{q}$ to obtain the estimated constellation symbols. The LLRs are extracted from the detected symbols to compute the achievable-rate loss, while the PAPR loss is calculated using the learned threshold parameter $\epsilon$, measured in dB. A detailed illustration of the full E2E learning pipeline is shown in Fig.~\ref{e2e arch}.

The neural architecture used to generate $\mathbf{Q}$ and $\mathbf{q}$ consists of a single complex-valued one-dimensional convolutional layer followed by batch normalization, which processes the channel impulse response. The resulting features are fed into a 1024-unit complex gated recurrent unit (GRU) layer and two fully connected layers. The final output is reshaped to produce the waveform matrix and detector parameters, which are then normalized according to (\ref{normalization factor}).

The modules responsible for generating the PAPR threshold $\epsilon_{\theta}$ and the uncertainty parameters employ lightweight fully connected neural networks composed of three real-valued dense layers with batch normalization. The learned PAPR threshold $\epsilon$ is constrained to the range $[2,8]$ dB, while the uncertainty parameters $\sigma_{\mathcal{R}},\sigma_{\mathcal{P}},\sigma_{\Theta}$ are clipped to the interval $[-10,10]$ to ensure stable training.

\subsection{Open-Source Implementation}

To encourage reproducible research and practical adoption, we release an open-source implementation of the DeepOFW framework. The implementation is built using the Sionna library~\cite{sionna}, which provides a flexible platform for modeling E2E communication systems and wireless channels. The source code is publicly available online~\cite{deepofw_code}, enabling researchers and practitioners to reproduce the results presented in this work and to explore extensions of DeepOFW for other wireless communication settings.

\section{Performance Evaluation}
\label{Performance Analysis}

In this section, we evaluate the performance of the proposed DeepOFW framework through numerical simulations. The proposed approach is compared with several well-known transmission schemes in order to assess its advantages in terms of PAPR and communication reliability under realistic wireless channel conditions.

\subsection{Simulation 1: PAPR Comparison}

We first examine the PAPR characteristics of DeepOFW. In this experiment, the proposed method is compared with conventional OFDM in order to highlight the improvement achieved by the learned waveform design.

We consider a system with $N=32$ subcarriers.
A 16-QAM modulation is used, forming a single time-domain symbol. For training, each batch sample consists of 384 QAM symbols per block, corresponding to 12 time-domain symbols. For evaluation, longer blocks consisting of 7,776 QAM symbols are used in order to obtain reliable PAPR statistics.

The wireless channel is modeled using the Tapped Delay Line (TDL) model of type 'A' according to the 3GPP specification~\cite{3gpp38901}. Each realization is generated with a uniformly sampled delay spread ranging from 10ns to 600ns, covering channel profiles from short indoor environments to large delay spreads such as the UMi street-canyon scenario~\cite{3gpp38901}. 
A baseband bandwidth of 1 MHz and a carrier frequency of 3.5 GHz are assumed. The SNR ranges from 0dB to 25dB, defined in terms of $E_b/N_0$. The noise variance is computed as $N_0 = E_s/(M E_b/N_0)$, where $E_s$ denotes the symbol energy.

During training, both the SNR and the channel delay spread are randomly sampled for each batch. A batch size of 9,000 and a learning rate of 0.001 are used. A second training stage is applied for fine tuning using shorter blocks consisting of a single time-domain symbol (32 QAM symbols), a batch size of 14,000, a learning rate of 0.0001, and higher SNR values ranging from 20dB to 25dB. The clipping range of $\epsilon$ is restricted to the interval of 2dB to 6dB.

\begin{figure}[h!]
\centering
\includegraphics[width=0.45\textwidth]{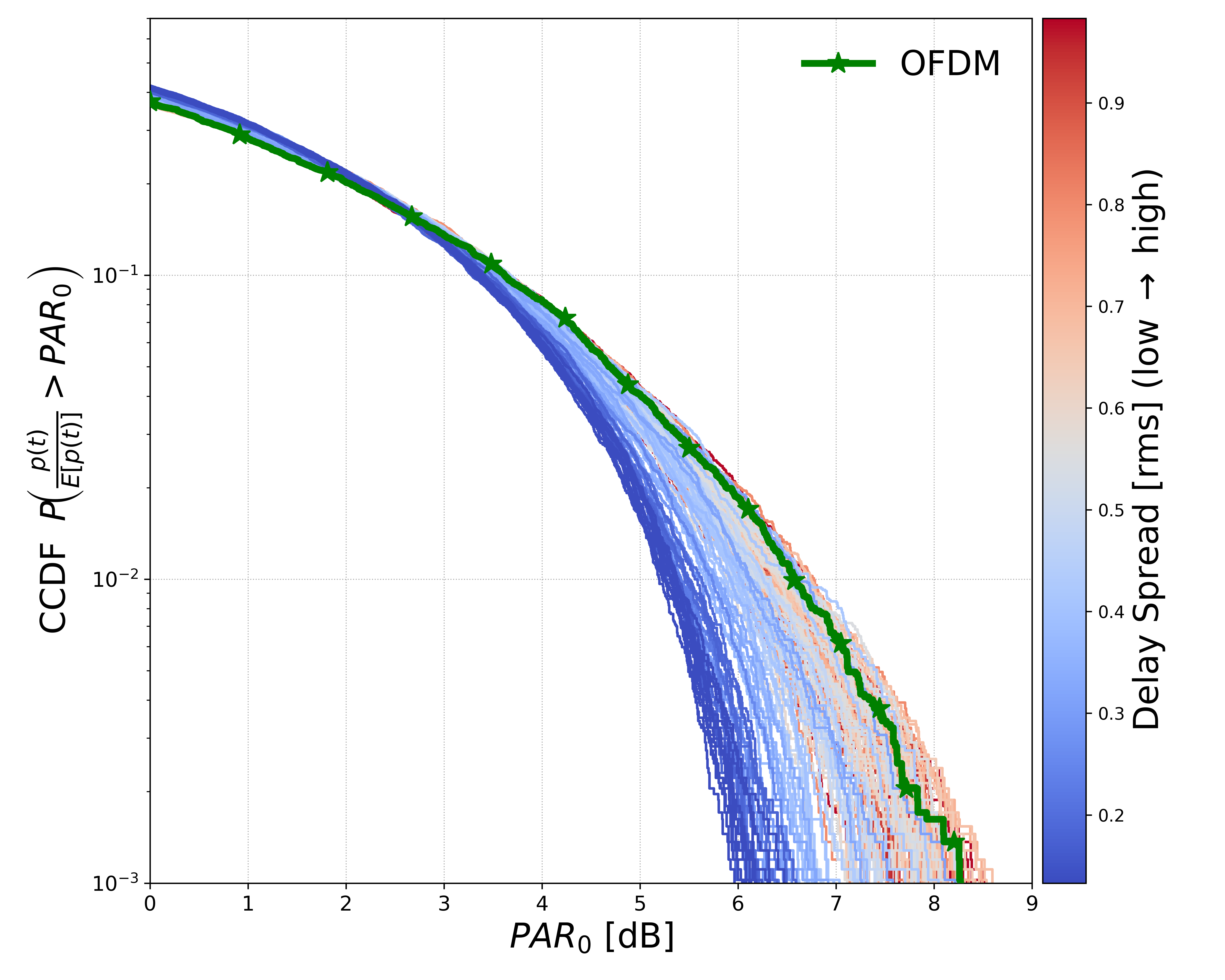}
\caption{CCDF comparison between conventional OFDM (green curve with star markers) and DeepOFW. Each DeepOFW curve corresponds to a different RMS delay spread realization, color-coded from blue (low delay spread) to red (high delay spread).}
\label{fig:CCDF}
\end{figure}

The complementary cumulative distribution function (CCDF) of the PAPR is presented in Fig.~\ref{fig:CCDF}. While the OFDM PAPR distribution is independent of the channel and therefore appears as a single reference curve, DeepOFW adapts its waveform representation to the channel conditions. Each curve corresponds to a different channel realization with a distinct RMS delay spread value. A total of 200 delay spread realizations were evaluated. The curves are color-coded according to the delay spread, ranging from blue (small delay spread) to red (large delay spread), illustrating how the learned waveforms adapt to the channel structure. The results demonstrate that DeepOFW achieves significantly improved PAPR characteristics compared with classical OFDM.

\subsection{Simulation 2: BER Comparison}

\begin{figure}[ht]
\centering
\includegraphics[width=0.5\textwidth]{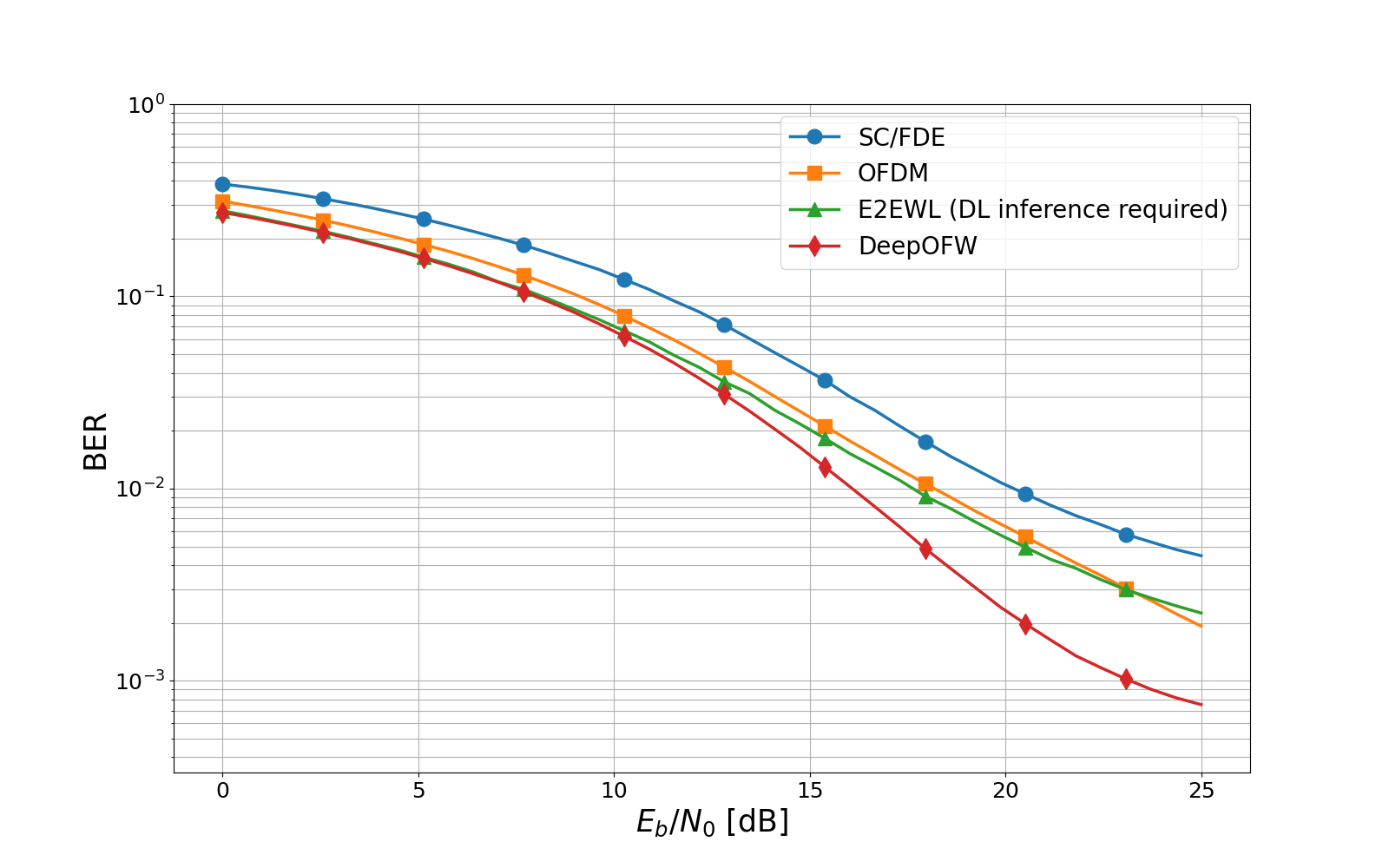}
\caption{BER performance of DeepOFW compared with conventional OFDM, SC/FDE, and E2EWL. Each point is averaged over 1,000 channel realizations with RMS delay spreads ranging from 10ns to 600ns.}
\label{fig:BER preformence}
\end{figure}

Next, we evaluate the communication reliability of the proposed approach. In this experiment, DeepOFW is compared with several well-known transmission schemes, including conventional OFDM, SC/FDE~\cite{al2008single}, and E2E waveform learning (E2EWL)~\cite{9747919}. SC/FDE~\cite{al2008single} is included as a classical baseline representing the single-user counterpart of the LTE uplink standard SC-FDMA. The SC/FDE implementation employs a Root-Raised-Cosine pulse-shaping filter with a roll-off factor of 0.15, an oversampling factor of 2, and a symbol span of 8. The E2EWL method~\cite{9747919} serves as a neural baseline. It is particularly relevant since it satisfies similar transmitter-side hardware constraints by restricting deep learning to the receiver side; however, unlike the proposed DeepOFW architecture, E2EWL relies on a high-complexity neural detector to achieve its performance. The training procedure of E2EWL follows the methodology of~\cite{9747919}, using $\epsilon_A = 0$ and $\epsilon_P = 8$ dB.

The BER performance obtained from this simulation is shown in Fig.~\ref{fig:BER preformence}. For each SNR value, the BER is computed by averaging over 1,000 channel realizations drawn from the same delay spread distribution used in the previous experiment. The results demonstrate that DeepOFW achieves significantly improved reliability compared with the considered baselines, while maintaining significantly lower receiver complexity than neural detection approaches such as E2EWL.

\subsection{Simulation 3: Learned Waveform Analysis}

\begin{figure}[H]
    \centering
    
    \begin{subfigure}[t]{0.45\textwidth}
        \centering
        \includegraphics[width=\textwidth]{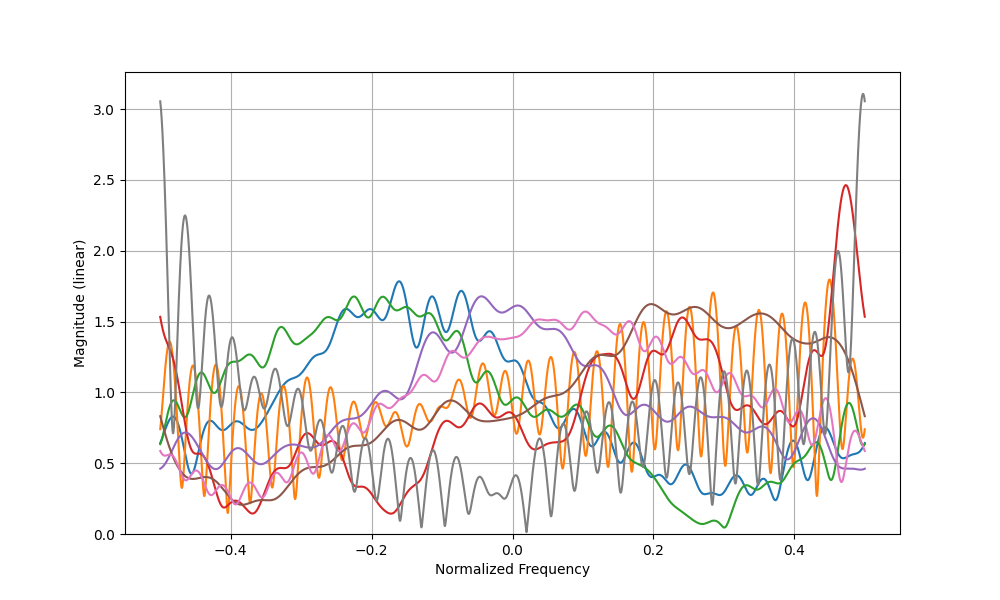}
        \caption{First eight waveform spectra for RMS delay spread of 10 ns.}
        \label{fig:freq_ds_10}
    \end{subfigure}\par\medskip

    \begin{subfigure}[t]{0.45\textwidth}
        \centering
        \includegraphics[width=\textwidth]{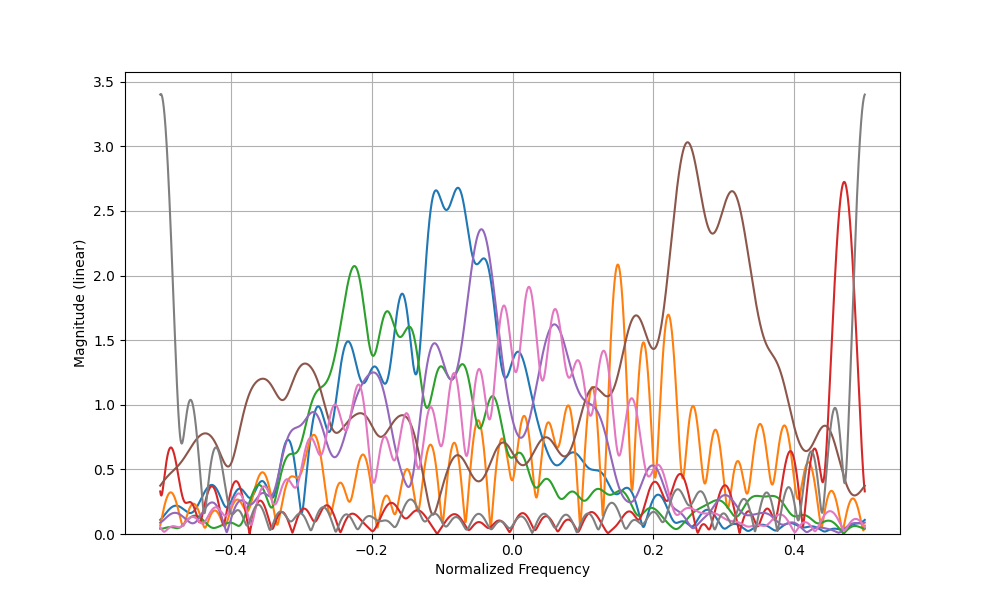}
        \caption{First eight waveform spectra for RMS delay spread of 130 ns.}
        \label{fig:freq_ds_130}
    \end{subfigure}\par\medskip

    \begin{subfigure}[t]{0.45\textwidth}
        \centering
        \includegraphics[width=\textwidth]{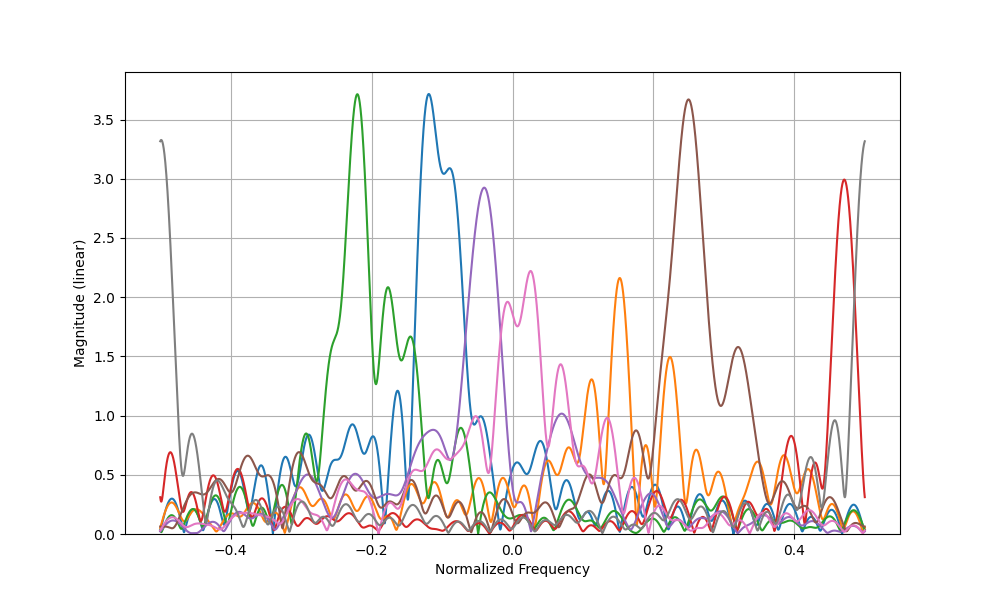}
        \caption{First eight waveform spectra for RMS delay spread of 250 ns.}
        \label{fig:freq_ds_250}
    \end{subfigure}\par\medskip

    \begin{subfigure}[t]{0.45\textwidth}
        \centering
        \includegraphics[width=\textwidth]{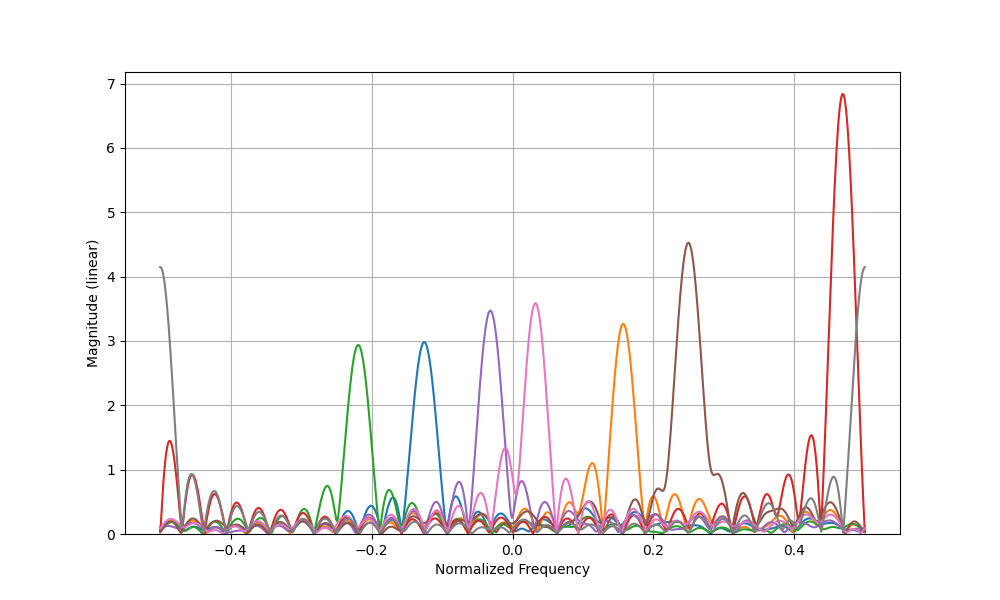}
        \caption{First eight waveform spectra for RMS delay spread of 580 ns.}
        \label{fig:freq_ds_580}
    \end{subfigure}\par\medskip    

    \caption{Frequency-domain representations of the first eight generated waveforms for different RMS delay spread conditions.}
    \label{fig:freq_waveforms_all_ds}
\end{figure}

\begin{figure}[!t]
\centering

\includegraphics[width=0.9\columnwidth]{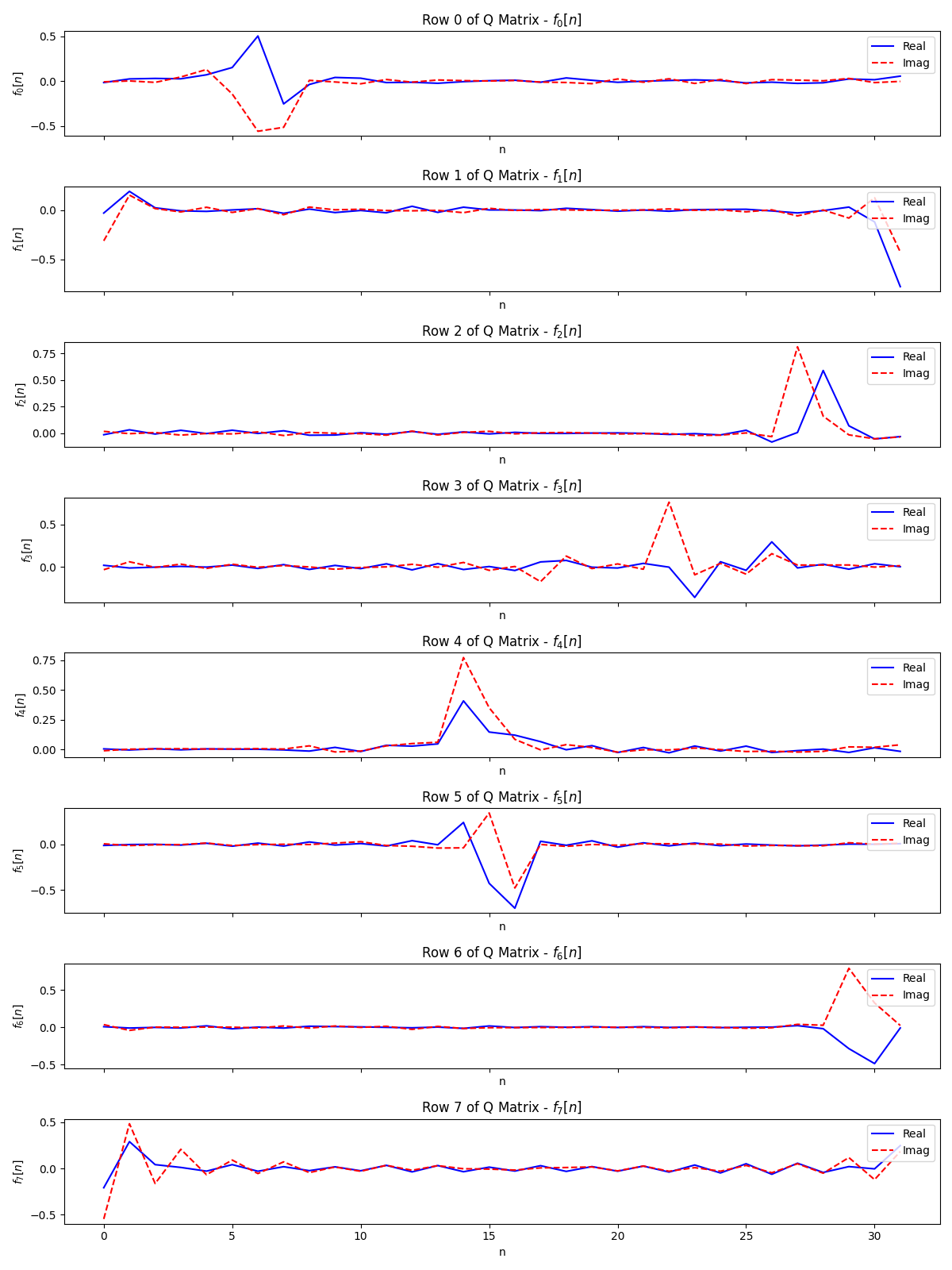}
\caption{Time-domain representations of the first eight DeepOFW-generated waveforms for an RMS delay spread of 10 ns.}
\label{fig:time_ds_10}

\vspace{2mm}

\includegraphics[width=0.9\columnwidth]{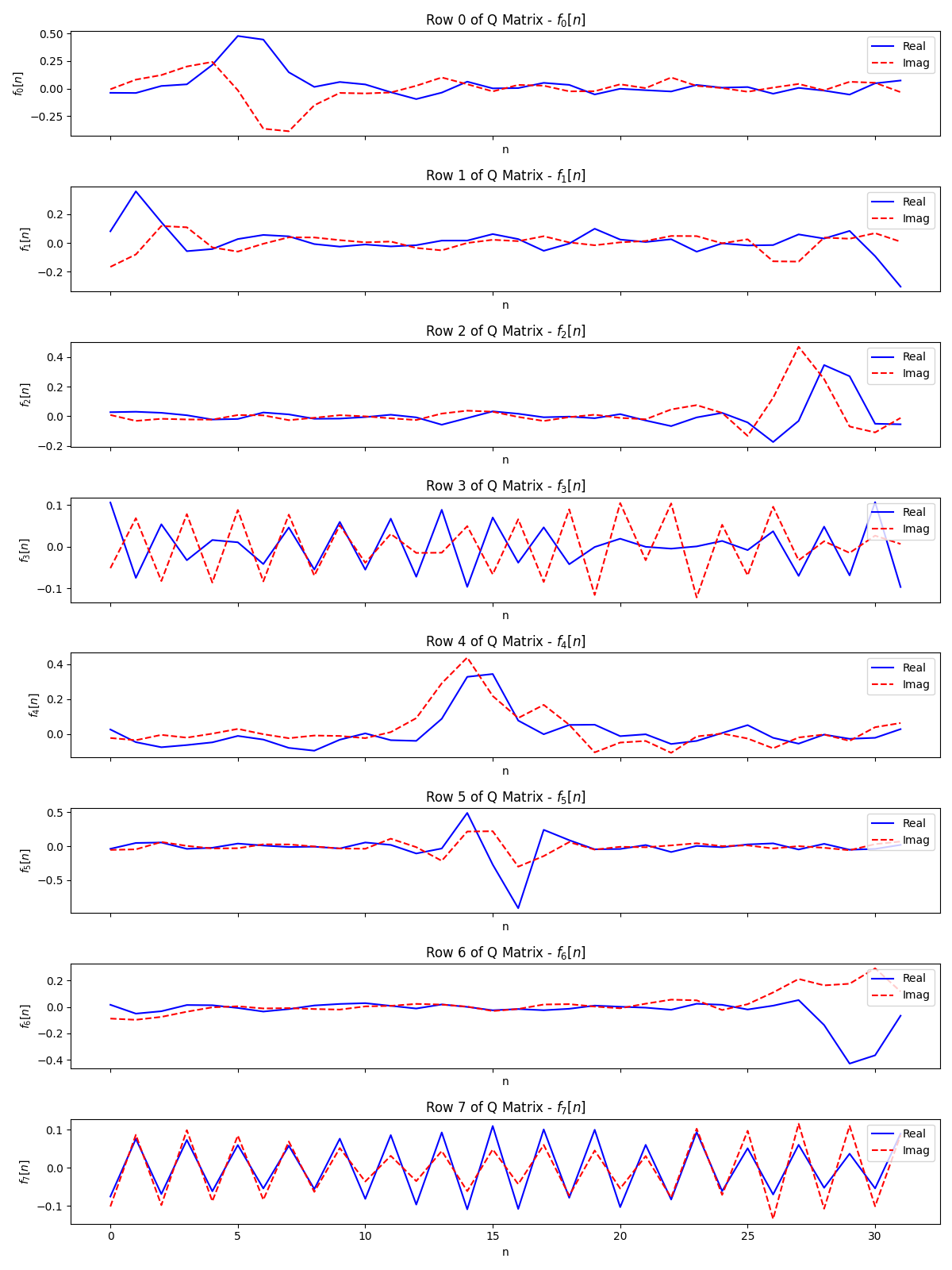}
\caption{Time-domain representations of the first eight DeepOFW-generated waveforms for an RMS delay spread of 130 ns.}
\label{fig:time_ds_130}

\end{figure}

\begin{figure}[!t]
\centering

\includegraphics[width=0.9\columnwidth]{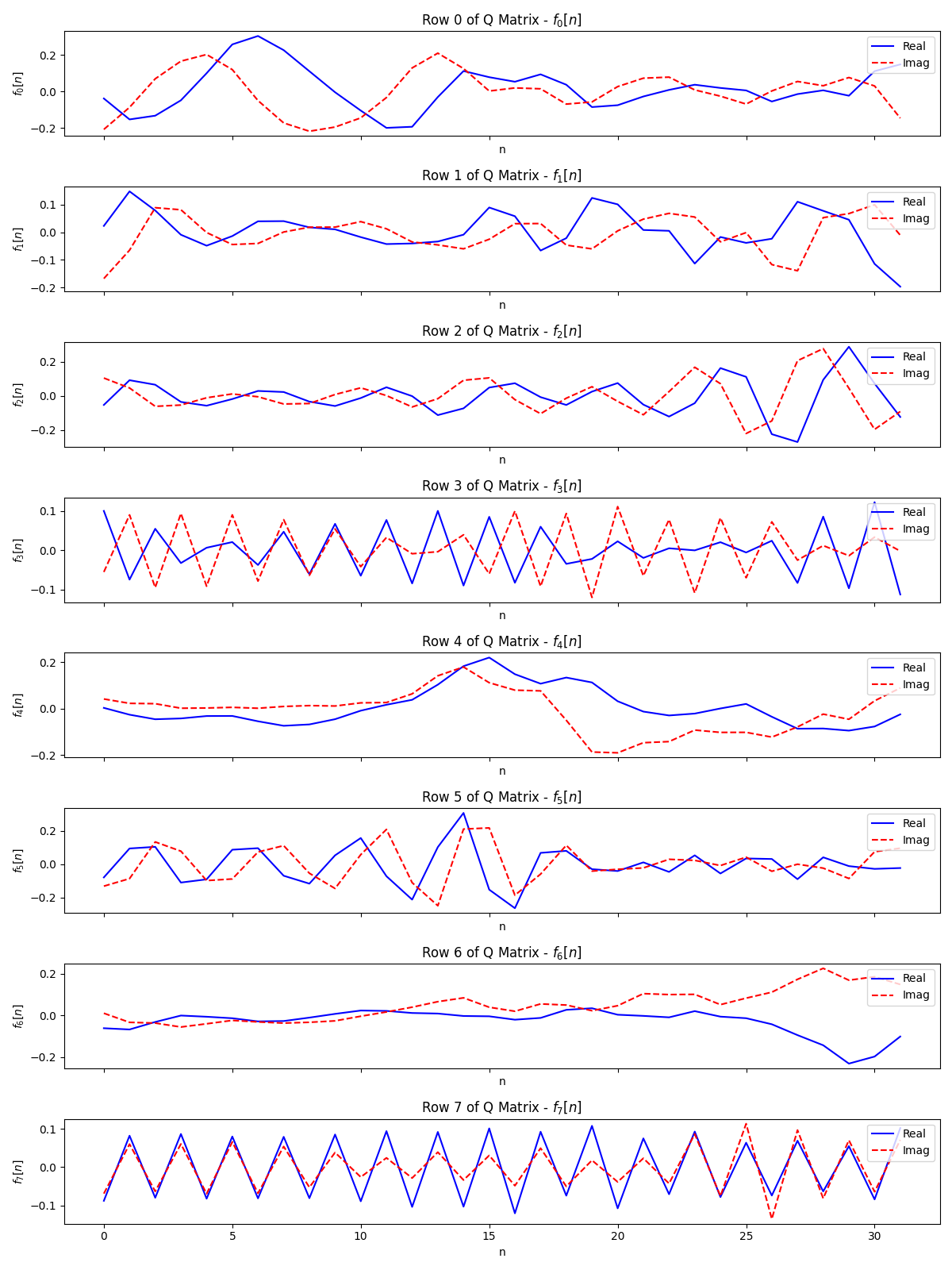}
\caption{Time-domain representations of the first eight DeepOFW-generated waveforms for an RMS delay spread of 250 ns.}
\label{fig:time_ds_250}

\vspace{2mm}

\includegraphics[width=0.9\columnwidth]{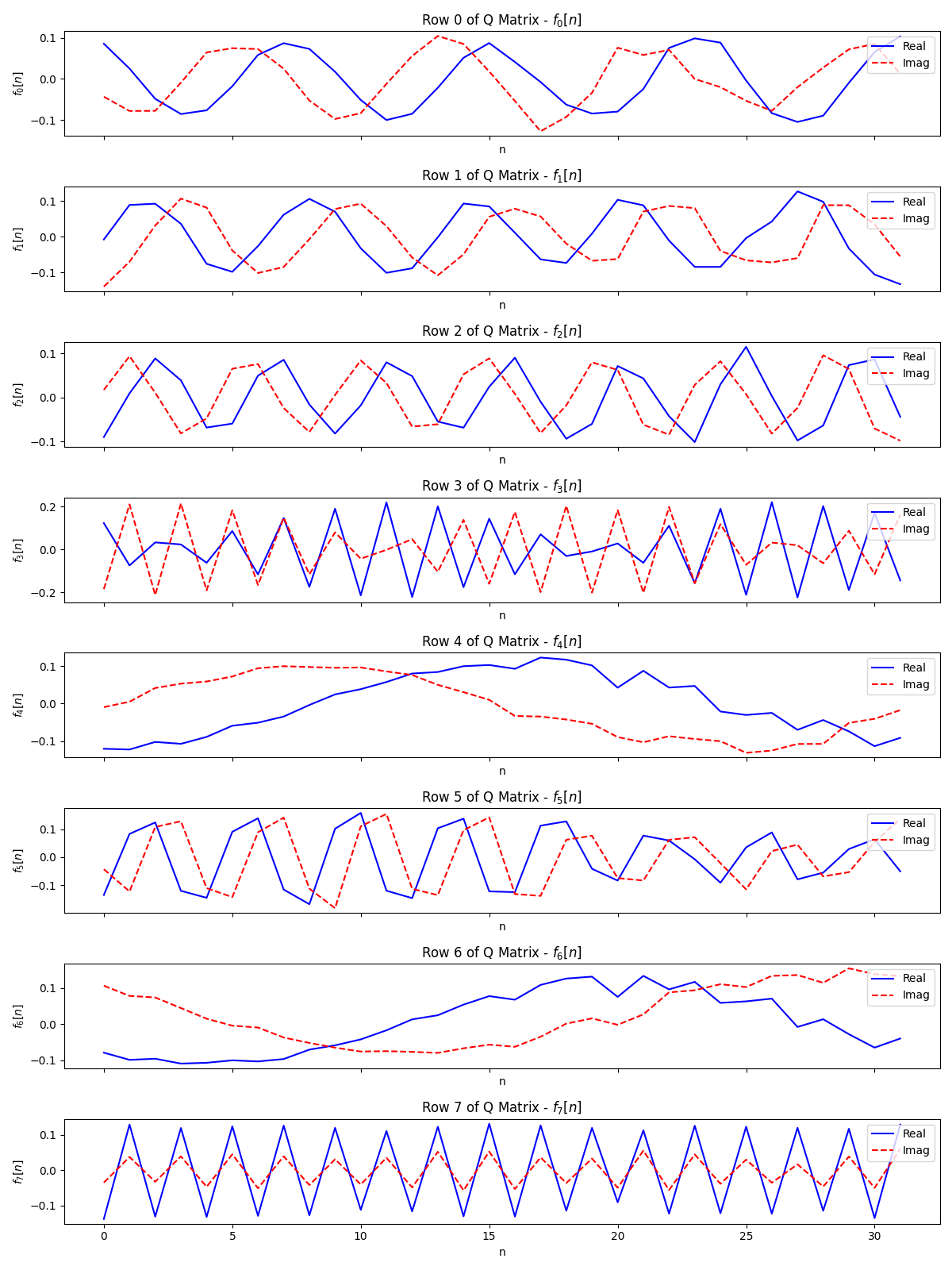}
\caption{Time-domain representations of the first eight DeepOFW-generated waveforms for an RMS delay spread of 580 ns.}
\label{fig:time_ds_580}

\end{figure}

To better understand how the proposed DeepOFW framework adapts its waveform structure to the wireless channel, we analyze the generated waveforms under different delay spread (DS) conditions. Four representative RMS delay spread values are considered: 10ns, 130ns, 250ns, and 580ns. For each case, the first eight generated waveforms are visualized in both the time and frequency domains in Figs.~\ref{fig:freq_waveforms_all_ds}, \ref{fig:time_ds_10}, \ref{fig:time_ds_130}, \ref{fig:time_ds_250}, 
\ref{fig:time_ds_580}.

The learned waveforms reveal a clear adaptation mechanism with respect to the channel profile. In particular, the model adjusts the structure of the waveform basis in order to mitigate large instantaneous peaks in the constructive summation of waveform components, which directly contributes to PAPR reduction.

For the low delay spread case (10ns), shown in Fig.~\ref{fig:freq_ds_10} and its corresponding time-domain representation in Fig.~\ref{fig:time_ds_10}, the model learns waveform structures whose summation resembles a serial transmission scheme. In this regime, the symbols are primarily separated in time, similar to a time-division multiplexing structure. This configuration naturally produces a very low PAPR, while reliable symbol recovery remains possible due to the relatively mild channel distortion.

As the delay spread increases, the model gradually adapts the waveform structure to cope with the more complex channel conditions. As illustrated in Fig.~\ref{fig:freq_ds_130}, Fig.~\ref{fig:freq_ds_250}, and Fig.~\ref{fig:freq_ds_580}, the learned waveforms become increasingly distributed across the frequency domain. This shift enables better mitigation of multipath interference and facilitates channel equalization, although at the cost of a moderate increase in PAPR.

In the highest delay spread scenario (580ns), the generated waveforms resemble scaled orthogonal complex sinusoids, approaching a structure similar to classical multicarrier modulation. Nevertheless, the resulting CCDF curves continue to outperform conventional OFDM waveforms while maintaining a lower error rate.

These observations reveal a key capability of the DeepOFW framework: the learned waveform representations dynamically adjust the time–frequency multiplexing structure according to the channel conditions. For channels with small delay spread, the model favors waveform structures that concentrate energy in time, producing low-PAPR signals that resemble serial transmission schemes. As the channel delay spread increases, the learned waveforms progressively distribute their energy across the frequency domain, enabling more effective mitigation of multipath interference while maintaining reliable detection. This adaptive behavior illustrates the ability of DeepOFW to automatically discover waveform structures that balance the trade-off between PAPR reduction and communication reliability. The results demonstrate that data-driven waveform design can flexibly tailor multicarrier modulation to the underlying channel characteristics, highlighting the strong potential of the DeepOFW framework as a powerful and hardware-efficient modulation paradigm for future wireless communication systems.

\vspace{-0.25cm}
\section{Conclusion}
\label{Conclusion}

This paper introduced DeepOFW, a deep learning–driven OFDM-flexible waveform modulation framework that enables data-driven waveform design while preserving the low-complexity hardware structure of conventional communication systems. By formulating the multicarrier transmission process as a fully differentiable architecture, the proposed approach allows E2E optimization of waveform generation and receiver processing under realistic physical constraints. Unlike existing neural transceiver designs, DeepOFW confines the deep learning inference stage to a centralized or offline environment, allowing the resulting waveform parameters to be deployed on standard transmitter and receiver hardware without additional computational overhead.

Simulation results over 3GPP multipath channel models demonstrate that the learned waveforms significantly reduce the PAPR compared with classical OFDM while simultaneously improving communication reliability relative to state-of-the-art transmission schemes. Furthermore, the learned waveform structures reveal an adaptive time–frequency multiplexing behavior that automatically adjusts to channel conditions in order to balance PAPR reduction and detection performance.

These results illustrate the strong potential of data-driven waveform design as a new paradigm for multicarrier communications. The proposed DeepOFW framework opens promising directions for future wireless systems, including energy-efficient transmitters, adaptive waveform design for next-generation cellular networks, and flexible modulation strategies for emerging communication scenarios such as dense wireless networks and intelligent radio environments.

\begingroup
\renewcommand{\baselinestretch}{1.0}
\normalsize

\bibliographystyle{IEEEtran}
\FloatBarrier
\bibliography{references}

@ARTICLE{8054694,
  author={O’Shea, Timothy and Hoydis, Jakob},
  journal={IEEE Transactions on Cognitive Communications and Networking}, 
  title={An Introduction to Deep Learning for the Physical Layer}, 
  year={2017},
  volume={3},
  number={4},
  pages={563-575},
  doi={10.1109/TCCN.2017.2758370}}

@techreport{3gpp38901,
  author={{3GPP Project}},  
  title        = {Study on Channel Model for Frequencies from 0.5 to 100 GHz},
  year         = {2020},
  url          = {https://www.3gpp.org/ftp/Specs/archive/38_series/38.901/},
}

@misc{sionna,
  author       = {J. Hoydis and S. Cammerer and F. Ait Aoudia and M. Nimier-David and
                  L. Maggi and G. Marcus and A. Vem and A. Keller},
  title        = {{Sionna}: An Open-Source Library for Next-Generation Physical Layer Research},
  year         = {2024},
  howpublished = {\url{https://nvlabs.github.io/sionna/}},
  note         = {Version 1.2.1}
}

@ARTICLE{9905727,
  author={Ozpoyraz, Burak and Dogukan, Ali Tugberk and Gevez, Yarkin and Altun, Ufuk and Basar, Ertugrul},
  journal={IEEE Open Journal of the Communications Society}, 
  title={Deep Learning-Aided 6G Wireless Networks: A Comprehensive Survey of Revolutionary PHY Architectures}, 
  year={2022},
  volume={3},
  number={},
  pages={1749-1809},
  doi={10.1109/OJCOMS.2022.3210648}}

@ARTICLE{10589475,
  author={Islam, Nazmul and Shin, Seokjoo},
  journal={IEEE Open Journal of the Communications Society}, 
  title={Deep Learning in Physical Layer: Review on Data Driven End-to-End Communication Systems and Their Enabling Semantic Applications}, 
  year={2024},
  volume={5},
  number={},
  pages={4207-4240},
  doi={10.1109/OJCOMS.2024.3425314}}

@ARTICLE{9446711,
  author={Zou, Cong and Yang, Fang and Song, Jian and Han, Zhu},
  journal={IEEE Communications Magazine}, 
  title={Channel Autoencoder for Wireless Communication: State of the Art, Challenges, and Trends}, 
  year={2021},
  number={5},
  pages={136-142},
  doi={10.1109/MCOM.001.2000804}}

@ARTICLE{10485272,
  author={Ye, Neng and Miao, Sirui and Pan, Jianxiong and Ouyang, Qiaolin and Li, Xiangming and Hou, Xiaolin},
  journal={IEEE Transactions on Cognitive Communications and Networking}, 
  title={Artificial Intelligence for Wireless Physical-Layer Technologies (AI4PHY): A Comprehensive Survey}, 
  year={2024},
  volume={10},
  number={3},
  pages={729-755},
  doi={10.1109/TCCN.2024.3382973}}

@ARTICLE{11053759,
  author={Doha, Shadman Rahman and Abdelhadi, Ahmed},
  journal={IEEE Access}, 
  title={Deep Learning in Wireless Communication Receivers: A Survey}, 
  year={2025},
  volume={13},
  number={},
  pages={113586-113605},
  doi={10.1109/ACCESS.2025.3584000}}

@ARTICLE{9271932,
  author={Luong, Thien Van and Ko, Youngwook and Matthaiou, Michail and Vien, Ngo Anh and Le, Minh-Tuan and Ngo, Vu-Duc},
  journal={IEEE Trans. on Wireless Communications}, 
  title={Deep Learning-Aided Multicarrier Systems}, 
  year={2021},
  volume={20},
  pages={2109-2119},
}

@ARTICLE{9747919,
  author={Ait Aoudia, Fayçal and Hoydis, Jakob},
  journal={IEEE Transactions on Communications}, 
  title={Waveform Learning for Next-Generation Wireless Communication Systems}, 
  year={2022},
  volume={70},
  number={6},
  pages={3804-3817},
  doi={10.1109/TCOMM.2022.3164060}}

@ARTICLE{9745361,
  author={Song, Jinxiang and Häger, Christian and Schröder, Jochen and Amat, Alexandre Graell I and Wymeersch, Henk},
  journal={IEEE Journal of Selected Topics in Quantum Electronics}, 
  title={Model-Based End-to-End Learning for WDM Systems With Transceiver Hardware Impairments}, 
  year={2022},
  volume={28},
  pages={1-14},
}

@book{9661102,
  author    = {Das, Suvra Sekhar and Prasad, Ramjee},
  title     = {Orthogonal Time Frequency Space Modulation: OTFS a Waveform for 6G},
  publisher = {River Publishers},
  year      = {2021},
  edition   = {1},
  doi       = {10.1201/9781003339021},
  url       = {https://doi.org/10.1201/9781003339021}
}

@article{Matthé2016,
  author    = {Matthé, M. and others},
  title     = {Precoded GFDM transceiver with low complexity time domain processing},
  journal   = {Wireless Networks},
  volume    = {138},
  year      = {2016},
  doi       = {10.1186/s13638-016-0633-1},
  url       = {https://doi.org/10.1186/s13638-016-0633-1},
  note      = {}
}

@INPROCEEDINGS{9013425,
  author={Omar, Muhammad Shahmeer and Ma, Xiaoli},
  booktitle={IEEE Global Communications Conference}, 
  title={Designing OCDM-Based Multi-User Transmissions}, 
  year={2019},
  volume={},
  number={},
  doi={10.1109/GLOBECOM38437.2019.9013425}}

@book{Holma2009,
  author    = {Holma, Heikki and T{\"o}monen, Antti},
  title     = {LTE for UMTS: OFDMA and SC-FDMA Based Radio},
  publisher = {Wiley},
  address   = {Hoboken, NJ, USA},
  year      = {2009},
  doi       = {10.1002/9780470745489},
  url       = {https://doi.org/10.1002/9780470745489}
}

@ARTICLE{9508784,
  author={Ait Aoudia, Fayçal and Hoydis, Jakob},
  journal={IEEE Transactions on Wireless Communications}, 
  title={End-to-End Learning for OFDM: From Neural Receivers to Pilotless Communication}, 
  year={2022},
  volume={21},
  number={2},
  pages={1049-1063},
  doi={10.1109/TWC.2021.3101364}}

@InProceedings{Kendall_2018_CVPR,
author = {Kendall, Alex and Gal, Yarin and Cipolla, Roberto},
title = {Multi-Task Learning Using Uncertainty to Weigh Losses for Scene Geometry and Semantics},
booktitle = {IEEE Conf. on Computer Vision and Pattern Rec. (CVPR)},
year = {2018}
}

@article{al2008single,
  title={Single-carrier frequency domain equalization},
  author={Al-Dhahir, Naofal and Uysal, Murat and Mheidat, Hakam},
  journal={IEEE Signal Processing Magazine},
  volume={25},
  number={5},
  pages={37--56},
  year={2008}
}

@article{juwono2021future,
  title={Future OFDM-based communication systems towards 6G and beyond: machine learning approaches},
  author={Juwono, Filbert H and Reine, Regina},
  journal={Green Intelligent Systems and App.},
  volume={1},
  pages={19--25},
  year={2021}
}

@article{shafie2024coexistence,
  title={On the coexistence of OTFS modulation with OFDM-based communication systems},
  author={Shafie, Akram and Yuan, Jinhong and Fitzpatrick, Paul and Sakurai, Taka and Fang, Yuting},
  journal={IEEE Trans. on Communications},
  pages={6822--6838},
  year={2024},
  publisher={IEEE}
}

@inproceedings{farhang2024sc,
  title={SC-FDMA as a delay-Doppler domain modulation technique},
  author={Farhang, Arman and Bayat, Mohsen},
  booktitle={IEEE International Conference on Communications (ICC)},
  pages={69--74},
  year={2024}
}

@article{tai2020overview,
  title={An overview of generalized frequency division multiplexing (GFDM)},
  author={Tai, Ching-Lun and Wang, Tzu-Han and Huang, Yu-Hua},
  journal={arXiv preprint arXiv:2008.08947},
  year={2020}
}

@article{moreira2024orthogonal,
  title={Orthogonal chirp-frequency division multiplexing},
  author={Moreira, T{\'u}lio Fernandes and de Lima Filomeno, Mateus and Martins, Wallace Alves and Ribeiro, Mois{\'e}s Vidal},
  journal={IEEE Trans. on Communications},
  pages={4630--4646},
  year={2024},
  publisher={IEEE}
}

@article{ouyang2016orthogonal,
  title={Orthogonal chirp division multiplexing},
  author={Ouyang, Xing and Zhao, Jian},
  journal={IEEE Trans. on Communications},
  volume={64},
  pages={3946--3957},
  year={2016}
}

@article{oshea2017introduction,
  author  = {Timothy J. O'Shea and Jakob Hoydis},
  title   = {An Introduction to Deep Learning for the Physical Layer},
  journal = {IEEE Transactions on Cognitive Communications and Networking},
  volume  = {3},
  number  = {4},
  pages   = {563--575},
  year    = {2017},
  month   = {Dec},
  doi     = {10.1109/TCCN.2017.2758370}
}

@inproceedings{aoudia2018end,
  author    = {Fares Aoudia and Jakob Hoydis},
  title     = {End-to-End Learning of Communications Systems Without a Channel Model},
  booktitle = {2018 52nd Asilomar Conference on Signals, Systems, and Computers},
  pages     = {298--303},
  year      = {2018},
  doi       = {10.1109/ACSSC.2018.8645360}
}

@article{qin2019deep,
  author  = {Zhaohui Qin and Huan Ye and Geoffrey Ye Li and Biing-Hwang Juang},
  title   = {Deep Learning in Physical Layer Communications},
  journal = {IEEE Wireless Communications},
  volume  = {26},
  number  = {2},
  pages   = {93--99},
  year    = {2019},
  month   = {Apr},
  doi     = {10.1109/MWC.2019.1800027}
}

@misc{deepofw_code,
  author       = {Ran Greidi and Kobi Cohen},
  title        = {{Open-Source Software for DeepOFW}},
  year         = {2026},
  howpublished = {\url{https://github.com/RanGreidi/DeepOFW}},
  note         = {GitHub repository}
}

@article{naparstek2012parametric,
  title={Parametric spectrum shaping for downstream spectrum management of digital subscriber lines},
  author={Naparstek, Oshri and Cohen, Kobi and Leshem, Amir},
  journal={IEEE communications letters},
  pages={417--419},
  year={2012},
  publisher={IEEE}
}

@article{leshem2012multichannel,
  title={Multichannel opportunistic carrier sensing for stable channel access control in cognitive radio systems},
  author={Leshem, Amir and Zehavi, Ephraim and Yaffe, Yoav},
  journal={IEEE J. on Selected Areas in Comm.},
  volume={30},
  number={1},
  pages={82--95},
  year={2012}
}

@article{ami2024stable,
  title={A Stable Polygamy Approach to Spectrum Access with Channel Reuse},
  author={Ami, Dan Ben and Cohen, Kobi},
  journal={arXiv:2408.12402},
  year={2024}
}

@ARTICLE{gafni2022distributed,
  author={Gafni, Tomer and Cohen, Kobi},
  journal={IEEE Access}, 
  title={Distributed Learning Over Markovian Fading Channels for Stable Spectrum Access}, 
  year={2022},
  volume={10},
  number={},
  pages={46652-46669},
  doi={10.1109/ACCESS.2022.3171666}}

@inproceedings{bistritz2018distributed,
  title={Distributed multi-player bandits-a game of thrones approach},
  author={Bistritz, Ilai and Leshem, Amir},
  booktitle={Advances in Neural Information Processing Systems},
  pages={7222--7232},
  year={2018}
}

@article{Tekin_2012_Online,
  title={Online learning of rested and restless bandits},
  author={Tekin, Cem and Liu, Mingyan},
  journal={IEEE Trans. on Information Theory},
  volume={58},
  number={8},
  pages={5588--5611},
  year={2012}
}

@article{liu2012learning,
  title={Learning in a changing world: Restless multiarmed bandit with unknown dynamics},
  author={Liu, Haoyang and Liu, Keqin and Zhao, Qing},
  journal={IEEE Transactions on Information Theory},
  volume={59},
  pages={1902--1916},
  year={2012},
  publisher={IEEE}
}

@ARTICLE{gafni2021learning,
  author={Gafni, Tomer and Cohen, Kobi},
  journal={IEEE Transactions on Automatic Control}, 
  title={Learning in Restless Multiarmed Bandits via Adaptive Arm Sequencing Rules}, 
  year={2021},
  volume={66},
  number={10},
  pages={5029-5036},
  doi={10.1109/TAC.2020.3043431}}

@inproceedings{cohen2014restless,
  title={Restless multi-armed bandits under time-varying activation constraints for dynamic spectrum access},
  author={Cohen, Kobi and Zhao, Qing and Scaglione, Anna},
  booktitle={Asilomar Conference on Signals, Systems and Computers},
  pages={1575--1578},
  year={2014}
}

@inproceedings{jiang2023online,
  title={Online restless bandits with unobserved states},
  author={Jiang, Bowen and Jiang, Bo and Li, Jian and Lin, Tao and Wang, Xinbing and Zhou, Chenghu},
  booktitle={International Conference on Machine Learning},
  pages={15041--15066},
  year={2023},
  organization={PMLR}
}

@article{raman2024global,
  title={Global rewards in restless multi-armed bandits},
  author={Raman, Naveen and Shi, Zheyuan R and Fang, Fei},
  journal={Advances in Neural Information Processing Systems},
  volume={37},
  pages={24625--24658},
  year={2024}
}

@article{naparstek2018deep,
  title={Deep multi-user reinforcement learning for distributed dynamic spectrum access},
  author={Naparstek, Oshri and Cohen, Kobi},
  journal={IEEE Transactions on Wireless Communications},
  volume={18},
  number={1},
  pages={310--323},
  year={2018}
}

@article{liu2021dynamic,
  title={Dynamic multichannel sensing in cognitive radio: Hierarchical reinforcement learning},
  author={Liu, Shuai and Wu, Jiayun and He, Jing},
  journal={IEEE Access},
  volume={9},
  pages={25473--25481},
  year={2021},
  publisher={IEEE}
}

@article{bokobza2023deep,
  title={Deep reinforcement learning for simultaneous sensing and channel access in cognitive networks},
  author={Bokobza, Yoel and Dabora, Ron and Cohen, Kobi},
  journal={IEEE Trans. on Wireless Comm.},
  pages={4930--4946},
  year={2023},
  publisher={IEEE}
}

@article{paul2023multi,
  title={Multi-Flow Transmission in Wireless Interference Networks: A Convergent Graph Learning Approach},
  author={Paul, Raz and Cohen, Kobi and Kedar, Gil},
  journal={IEEE Transactions on Wireless Communications},
  year={2023},
  publisher={IEEE}
}

@article{cohen2024sinr,
  title={SINR-aware deep reinforcement learning for distributed dynamic channel allocation in cognitive interference networks},
  author={Cohen, Yaniv and Gafni, Tomer and Greenberg, Ronen and Cohen, Kobi},
  journal={IEEE Transactions on Wireless Communications},
  year={2024},
  publisher={IEEE}
}

\endgroup


\end{document}